\documentclass[aps,prd,reprint,groupedaddress,nofootinbib]{revtex4-1}

\usepackage{amsmath}
\usepackage{amssymb}
\usepackage{graphicx}
\usepackage{epstopdf}
\usepackage[caption=false]{subfig}

\begin{document}


\title{Toy model studies of tuning and typicality with an eye toward cosmology}


\author{Aaron Hernley}
\author{Andreas Albrecht}
\affiliation{University of California at Davis;
Department Of Physics\\
One Shields Avenue;
Davis, CA 95616\\
}
\author{Tevian Dray}
\affiliation{
Department of Mathematics\\
Oregon State University\\
Corvallis, OR  97331
}


\date{\today}

\begin{abstract}
We investigate a number of simple toy models to explore interesting
relationships between dynamics and typicality. We start with an
infinite model that has been proposed as an illustration of how
non-ergodic dynamics can produce interesting features that are
suggestive for cosmological applications.  We consider various
attempts to define the infinite model more rigorously as a limit of a
finite system. None of our attempts at such rigor were able to preserve the
attractive properties.  We hope our work will challenge others to find
more successful toy models.  The difficulty of finding such models
suggests that connections between dynamics and typicality we hope for
in cosmological theories such as eternal inflation may not be so easy to achieve.
\end{abstract}


\maketitle

\section{Introduction}

Since the introduction of the idea of cosmic inflation 
 \cite{Starobinsky:1980te,Guth:1980zm,Linde:1981mu,Albrecht:1982wi} 
many
cosmologists have believed that a full theory of the cosmos should
give us reason to think that the universe we observe is typical in some
sense. This goal has proved difficult to achieve.  The popular eternal
inflation picture%
\footnote{For an excellent survey of this topic see
 \protect\cite{Guth:2007ng}}
currently struggles under the measure problem as
well as Page's Born rule problem (probably the deeper of the
two \cite{Albrecht:2012zp}).  The competing ``de Sitter Equilibrium''
picture \cite{Albrecht:2009vr,Albrecht:2011yg} requires a controversial
completion of the low 
energy effective theory, although it does avoid the measure and Born
Rule problems and does exhibit typicality of the observed universe.
The cyclic model, which is often offered as an alternative, has yet to
be analyzed systematically regarding the 
typicality of our universe.  For an overview of some of these issues see 
 \cite{Albrecht:2012aa}. 

This paper seeks to take an incremental step in the study of
these challenging issues by examining a number of toy models.  The
goal of these studies is to ask what can be said about ``typicality''
in some simple systems.  We do not claim to offer grand insights into
the cosmological problems here, as the toy models we consider are far
simpler than realistic cosmologies.   We only comment briefly about
how certain features of our toy models might inform our thinking about
cosmology.  Still, we have found this investigation interesting and we
believe this research can help us work toward more disciplined
thinking about the cosmological case. 

The starting point for our work is a toy model proposed by
Guth \cite{Guth:2011aa} which is explicitly non-ergodic and has some very
interesting features regarding typicality. We present this model in
Section \ref{Sec:USHOinfinite}. This toy model has
traditionally been presented as a formally infinite system. We next
study the same toy model where its infinite nature is more systematically
constructed as a limit of a finite case (Sections
\ref{Sec:USHOfinite} and \ref{Sec:USHOtrajectories}).
We show that the properties identified in the
formally infinite case only appear if the limit is taken in a
particular way, a way which requires constraints to be placed on the initial
conditions. From that point of view the behavior originally identified
appears to be less generic. 

We next take up the question of whether a different toy model might
exhibit significantly different properties (Section \ref{Sec:Bent}).
We investigate a second
toy model with the phase space distorted or ``bent'' relative to that
of the original model.  The specific goal is to achieve an outcome
different from that of the original model.  Our analysis shows
however that the bent model does not behave significantly differently
from the original model, and if anything exhibits less of the desired
behavior.

We present our interpretation and conclusions in Section
\ref{Sec:Conclusions}. Our work raises concerns that the properties
identified by Guth in his original toy model may not offer a path
toward understanding typicality in cosmology, at least not without
adding assumptions about initial conditions.

\section{Heuristic Description of the Infinite Case}
\label{Sec:USHOinfinite}
\par
\begin{figure}
\includegraphics[scale = 0.4]{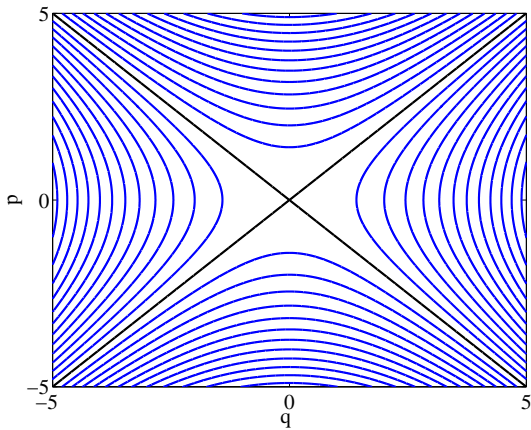}
\caption{Trajectories for the Upside-down Simple Harmonic Oscillator
  (USHO). The left and right quadrants correspond to bound states,  
whereas the top and bottom quadrants correspond to $E > 0$. Squeezing
toward the two asymptotes is an apparently universal behavior of this system.
\label{InfTrajFig}}
\end{figure}

Alan Guth has proposed a simple toy model that shows a natural
tendency toward fine-tuning of
parameters in an infinite phase space \cite{Guth:2011aa}. The Hamiltonian
for this model, 
\begin{equation}
H_{PQ} = \tan^{-1}{(-PQ)}, \label{guthham}
\end{equation}
is a canonical transformation of an upside-down simple harmonic
oscillator (USHO), a particle moving in the potential 
\begin{equation}
V(q) = -\frac{q^2}{2}. \label{infpot}
\end{equation}
Throughout this paper we use reduced coordinates which lead to the
simple equation of motion
\begin{equation}
\ddot{q} = -\frac{dV}{dq}.
\label{Eqn:OfMotion}
\end{equation}
We consider the USHO model and we see that there are four quadrants in
phase space (Fig.~\ref{InfTrajFig}). The left quadrant corresponds to
negative energy states where the particle moves right from $q =
-\infty$, reaches its turning point, and returns to $-\infty$. The
right quadrant corresponds to the mirror image of this where the
particle starts at $q = +\infty$, reaches its turning point, and
returns to $+\infty$. The top quadrant corresponds to the positive
energy states where the particle starts at $q = -\infty$, continues
over the top of the potential, and moves toward $+\infty$. The bottom
quadrant is the mirror image of the top quadrant where the particle
starts at $q = +\infty$, continues over the top of the potential, and
moves toward $-\infty$. 

\par
The key feature of this model is the flow of all trajectories toward
the two diagonal asymptotes.  This feature, known as ``squeezing'',
appears equivalently in each quadrant and for concreteness we zero in on
the quarter of phase space that corresponds to bound states 
($E < 0$) where $q > 0$. In this region, the trajectories are squeezed
along the line $p = -q$ in the infinite past and along the line $p =
q$ in the infinite future. 

The special typicality property of this system can be stated as
follows: Given any normalizable probability
distribution, and some minimum amount of squeezing $S$ and small
tolerance $\epsilon$ there will be some time, $T$, such that for every $t >
T$, the probability of finding the system squeezed at least as much as
$S$ will be greater than $1-\epsilon$. In other words, this system will spend
an infinite amount of time in a squeezed state.  We will discuss how
to quantify squeezing in the next section, but the essence of this point can
be readily 
seen by inspecting Fig.~\ref{InfTrajFig}.  Represent any normalizable
probability distribution by the contour of your choice (say, the one
that encloses $99\%$ of the probability). Then follow the evolution of
that distribution by allowing each point on the contour to move along
the trajectory on which it resides.   Clearly the contour will get
squeezed down to any set tolerance within a finite amount of time, and
then spend the rest of eternity squeezing even further than that. 

This is certainly not an ergodic system.  If you inspect the system
at a random time, it is highly unlikely to be found in just about
any part of phase space, and is most likely to be found in a state
arbitrarily squeezed against one of the asymptotes.  The system is
``naturally fine tuned'', in that a very special part of phase space
(as measured by phase space area) is strongly preferred dynamically.
One might hope that some behavior of this general sort might help us
show how the finely tuned initial conditions of the big bang (or the
apparently even more finely tuned initial conditions for an early
inflationary phase) could be dynamically favored.

More realistic models might include a second phase of evolution
during which the ``attractor'' region of phase space would be ``on
top of a hill'' or dynamically interesting in some other way, or
might distinguish between different past and future asymptotic
behaviors.  It is far from clear that the lessons we extract from
the toy model considered here will survive if the complexity is
increased in this way.  We nonetheless hope that starting with the
simplest toy models may be helpful in achieving a better
understanding of these difficult problems.  In particular, we treat
squeezing against either asymptote as equivalent, since at the level
of our simple models they are interchangable behaviors by a simple
$t \rightarrow -t$ transformation.

Our discussion so far has dealt with infinities in a rather
informal way. A more rigorous approach would be to consider this
system as the limit of some finite system as it becomes very large.
We do just that in the next section, and show that the limit must be
taken in a special way to allow the interesting dynamical fine tuning
to emerge. 


\section{A Finite Case}
\label{Sec:USHOfinite}
\par
\begin{figure}
\includegraphics[scale = 0.4]{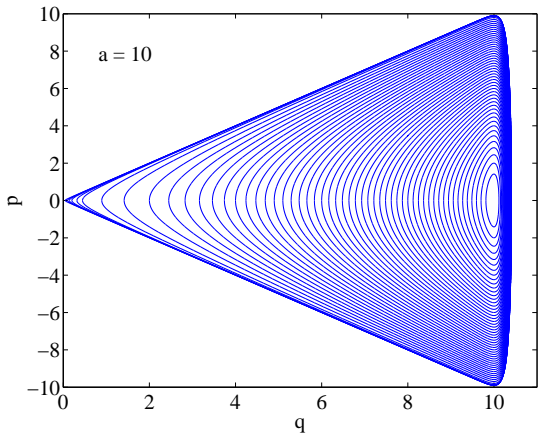}
\caption{Trajectories for the USHOb system with the barrier placed at
  $a = 10$. Introducing the barrier makes the system finite, creating
  a regulated version of the USHO system. 
(We show only one quadrant of phase space where $E < 0$ and $q
> 0$). \label{WallTrajFig}} 
\end{figure}

We consider a simple way of making the upside-down simple harmonic
oscillator finite by putting a sharp barrier at $q = a$, where $a$ is chosen
based on how large we want our system. To achieve this we add a
second term to the potential so that it becomes  
\begin{equation}
V(q) = -\frac{q^2}{2}+\left(\frac{q}{a}\right)^{100}. \label{wallpot}
\end{equation}
We call this the ``USHO with barrier model'' (USHOb). Once again, we focus on
the $E < 0$ and $q > 0$ quadrant. Trajectories are shown in
Fig.{\ref{WallTrajFig}}. By increasing $a$, we increase the size of
our system. Because the barrier is so sharp, most of the phase space
to the left of the barrier is unchanged from the infinite case. But
when the system approaches the barrier it rapidly reverses direction,
sending the trajectory rapidly down the plot to connect with a
trajectory with values of $p$ which are the opposite to the values
before the barrier collision.  

We consider the question: ``Can the infinite case be
thought of as the limit of this finite system as $a \rightarrow
\infty$?'' Because the trajectories in phase space are now closed,
we can no longer say that a given trajectory will spend an infinite (or
even necessarily an ever-increasing) amount of time in a squeezed
state. In fact, there are some trajectories that will never enter a
squeezed state. We now attempt to quantify these statements. 

\par
Begin by creating a patch of points in some region of phase
space. This patch has a well-defined area in phase space and as it
evolves in time along these trajectories, the phase space area will be
conserved due to Liouville's Theorem. This means that as it enters a
region of larger squeezing, the patch is stretched along one direction
and is squeezed along another. As an example, a circular patch
starting in a region near $p = 0$ for the infinite USHO model will
become more stretched out along the direction parallel to $p = q$ and
will become more squeezed along the direction perpendicular to $p =
q$. The circular patch becomes more and more elliptical as time passes
but the total area of this patch will remain constant. A problem with
defining a squeeze parameter on this patch is that for a given
starting position 
of the circular patch, the amount of squeezing will be dependent on
the size of the patch. One way around this is to introduce a local
definition for a squeeze parameter that will measure the amount of
squeezing at a given point in phase space.

We choose to quantify squeezing in terms of the extent to which a
given point is closer to one asymptote than the other.  Even though
``distance'' has no physical meaning in phase space, the ratios of
the distances to the two asymptotes are largely independent of the
details of how those distances are computed, and to some extent even
whether the geometry is assumed to be Euclidean.  In any case,
shifting our focus from the shape of the distribution to the
proximity of the asymptotes allows us to construct a more useful
squeeze parameter.

\par
The (shortest, Euclidean) distance between a given point in
phase space and the asymptote with slope $+1$ is given by
$d_+=|{p-q}|/\sqrt{2}$, and the distance to the other asymptote is
given by $d_-=|{p+q}|/\sqrt{2}$. We define our local squeeze
parameter by
\begin{equation}
s \equiv \frac{d_+}{d_-} + \frac{d_-}{d_+}. \label{sDef}
\end{equation}
By this definition, the squeeze parameter is large whenever the phase
space point is much closer to one asymptote than the other (regardless
of which one is closer). Note that this is not at all a generic
definition of squeezing, but instead makes use of the features of this
particular model that cause the trajectories to be squeezed against
particular asymptotes specific to the model.

\begin{figure}
\includegraphics[scale = 0.47]{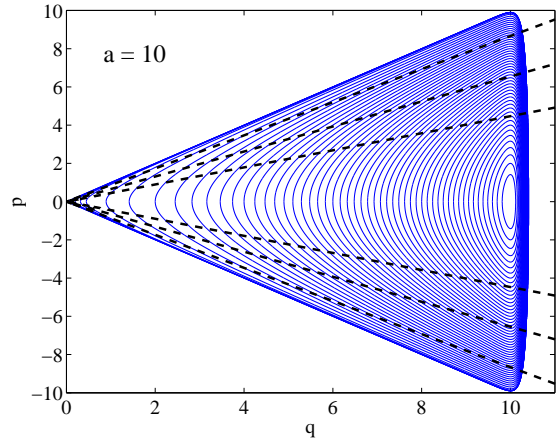}
\caption{The same phase space plot as Fig.~\ref{WallTrajFig}, with
  lines of constant squeeze parameter $S$ shown superposed (dashed) on
  top of the USHOb trajectories. From top
  to bottom, $s = 14, 5, 3, 3, 5, 14$.  \label{WallTrajFigCS}}
\end{figure}
Figure \ref{WallTrajFigCS} is equivalent to Fig.~\ref{WallTrajFig}
with the addition of 
a few constant $s$ lines (solving Eqn.~\ref{sDef} for $p(q)$ with
constant $s$ gives a pair of lines). The linear behavior of constant $s$ curves 
contrasted with the flow of the phase space trajectories
strongly toward the asymptotes may suggest that our squeeze
parameter is rather generously defined to favor large values. However,
since this paper ultimately argues that squeezing is not as
generically easy to achieve as one might have thought, making a
generous definition of squeezing helps us make a more robust argument. 


\section{Trajectories}
\label{Sec:USHOtrajectories}
\par
It is a simple task
to lay down a set of trajectories for a given placement of the
barrier.  What fraction of time is spent above a given squeeze
threshold by a given trajectory?  By a given \textit{distribution}
of trajectories?  What fraction of the \textit{area} of phase space
is above a given squeeze threshold?  We discuss each of these
questions in turn, in each case considering the limit as the squeeze
threshold $a$ approaches infinity.
 
\par
\begin{figure}
\includegraphics[scale = 0.4]{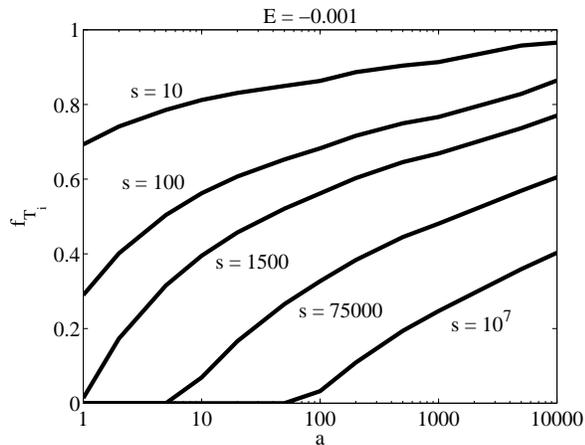}
\caption{Fraction of time squeezed $f_{T_{i}}$ vs. barrier position $a$ for
  selected 
  squeeze thresholds for an energy, $E = -0.001$, favorable to
  squeezing. The squeeze fraction increases as the barrier is moved
  to larger $a$ values. \label{WallE1e3TimeFig}} 
\end{figure}

We start by selecting a large energy trajectory, since such
trajectories will experience a lot of squeezing. 
The fraction of time a trajectory is squeezed is given by
\begin{equation}
f_{T_i} \equiv \frac{\Delta t_{\textrm{squeezed}}}{T_{\textrm{total}}}.
\label{sqonetraj}
\end{equation}
where $\Delta t_{\textrm{squeezed}}$ is the time spent above a given
squeeze threshold and $T_{\textrm{total}}$ is the total time for the
trajectory.  Several squeeze thresholds are shown in
Fig.~\ref{WallE1e3TimeFig},
with $E=-0.001$, very close to the upper bound of $E=0$ for this quadrant.
As we increase the size of the system by moving out the barrier, the
fraction of time that this trajectory spends squeezed is continually
increasing. This corresponds to the point of view described in
Section \ref{Sec:USHOinfinite}.

\par
\begin{figure}
\includegraphics[scale = 0.55]{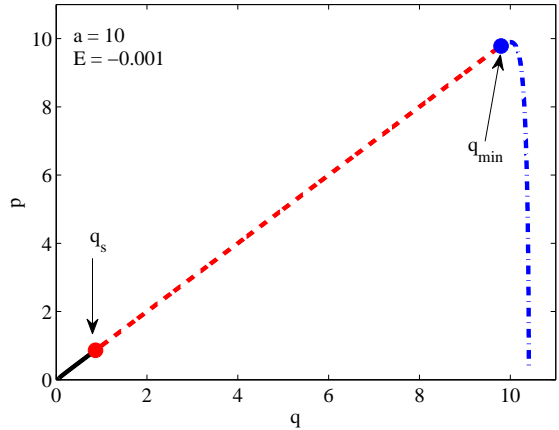}
\caption{Regions used for our analytical treatment of fraction of
time squeezed, with a threshold of $s=1500$, shown only for the half
of the trajectory with $p>0$.  The dashed line in the middle shows
the squeezed portion of the trajectory.}
\label{WallE1e3anal}
\end{figure}

 For an energy favorable to squeezing such as the one selected,
we can derive this behavior analytically. Due to the symmetry across
the $p=0$ axis, we only look at $p>0$, and we break the trajectory
into three distinct portions, as shown in
Fig.~\ref{WallE1e3anal}. The first goes from $p = 0$ to $p=q_s$, the
position where $s$ becomes greater than a given value. For
sufficient squeezing, this point is independent of $a$. This portion
of the trajectory is unsqueezed, and the time spent here ($\Delta
t_1 \equiv t_s - t_0$) is a constant, $c_1$, as the size of the
system increases.  
(The linear appearance of the trajectory for small $q$ is due to its
highly squeezed nature; the deviations from squeezing for $q<q_s$
are so small as not to be visible on the plot.)
The second portion of this trajectory, $p=q_s$ to
$p=q_{\textrm{min}}$ is squeezed, and extends to the point where $s$
again falls below the threshold value, which occurs at approximately
the minimum of $V$ and is dependent on the value of $a$.  Since the
system is highly squeezed, the time $\Delta t_2 = \int dq/p$ spent
on this portion can be approximated by $\int dq/q$, and $\Delta t_2$
increases logarithmically with barrier position $a$.  The third
portion is once again unsqueezed, and runs from $p=q_{\textrm{min}}$
to the turning point of the potential.  The time $\Delta t_3$ spent
here is constant as well, which we call $c_3$. Referencing
Eqn.~(\ref{sqonetraj}), we see
\begin{equation}
f_{T_i} \approx
\frac{\ln{\left({a}\right)}}{c_1+c_3+\ln{\left({a}\right)}}
\label{ftian}, 
\end{equation}
which goes to $1-\epsilon$ with $\epsilon \equiv (c_1+c_3)/\ln(a)$ for large
values of $a$.

\par
Of course this is only one trajectory, and one more favorable to
squeezing.  Next we consider a discrete set of trajectories,
covering the entire quarter of phase space, and average over them.
However, the choice of trajectories affects the final results.
To make an analogy to a ball rolling down a hill, each energy
corresponds to the initial placement of the ball, and choosing more
trajectories of larger energy means selecting more energies near the
top of the hill.  Since these trajectories become more squeezed than
those of smaller energies, there should also be a bias to the
fraction of time squeezed that is dependent on the set of
trajectories we chose.

\par
We choose a distribution of trajectories evenly spaced in energy up to
$E = -10$. For a given distribution of trajectories such as this, we
define the 
fraction of time this distribution spends squeezed as the simple
average
\begin{equation}
f_{T} \equiv \frac{1}{N_T}\sum{f_{T_i}}, \label{sqtraj}
\end{equation}
where $f_{T_i}$ is the fraction of time a given trajectory spends
squeezed and $N_T$ is the number of trajectories. Once again, we
arbitrarily choose several squeeze thresholds.

\begin{figure}
\includegraphics[scale = 0.54]{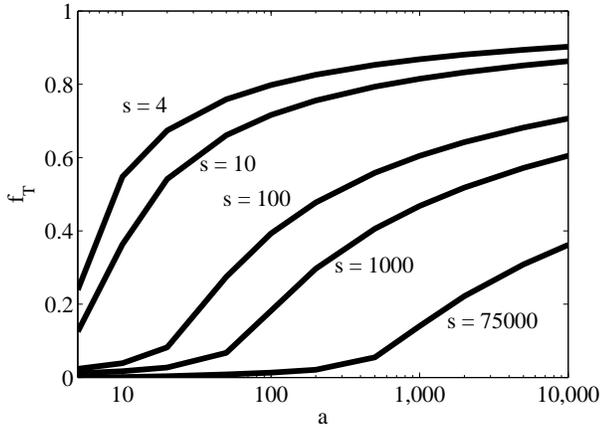}
\caption{Fraction of time squeezed $f_{T}$ vs. $a$ for selected
  squeeze thresholds evaluated over an entire family of trajectories
  biased toward energies that favor squeezing.
\label{WallTTimeFig}} 
\end{figure}

\begin{figure}
\includegraphics[scale = 0.54]{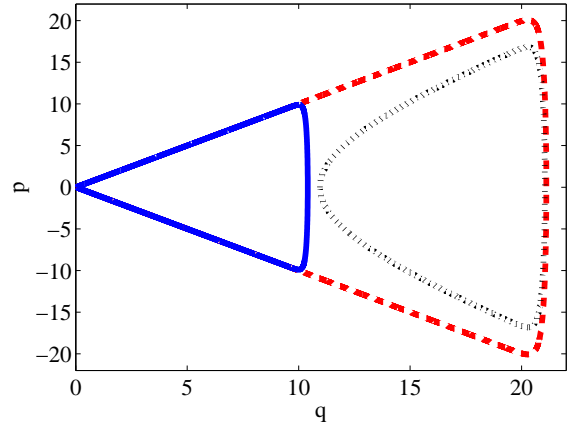}
\caption{Phase space trajectories illustrating how moving the barrier
  to larger $q$ values allows additional low energy trajectories that
  are less squeezed. The solid and dashed curves show trajectories for the
  same energy value,
but the dashed curve has the barrier moved out to
$a=20$ (vs $a=10$ for the solid curve).  The dotted curve is an example
of a trajectory with an energy that is forbidden when $a=10$, but is
allowed when $a=20$ (or larger). Clearly the dotted curve is less
squeezed than the others.
\label{TwoSizeTrajFig}}
\end{figure}

As shown in Fig.~\ref{WallTTimeFig}, as we increase the size of the
system by moving out the barrier (while keeping the trajectory
energies fixed), we again see that the fraction of time each
trajectory spends squeezed continues to increase, and thus so does
the average.  As we move the barrier out there is an ever-increasing
region of phase space that never reaches a fixed amount of squeezing
(corresponding to the large central region of unsqueezed phase space
clearly visible in Fig.~\ref{WallTrajFig} and illustrated in
Fig.~\ref{TwoSizeTrajFig}). But because we have fixed the trajectory
energies to correspond to squeezing even for close barrier
positions, the growing region of unsqueezed trajectories is not
probed by our distribution. Thus, the limit we take here matches the
intuition described in Sect.~\ref{Sec:USHOinfinite}, but does so at
least in part because the distribution of trajectories is biased
toward those that are more squeezed.

\par
\begin{figure*}
\centering
\subfloat[]{
\includegraphics[scale = 0.41]{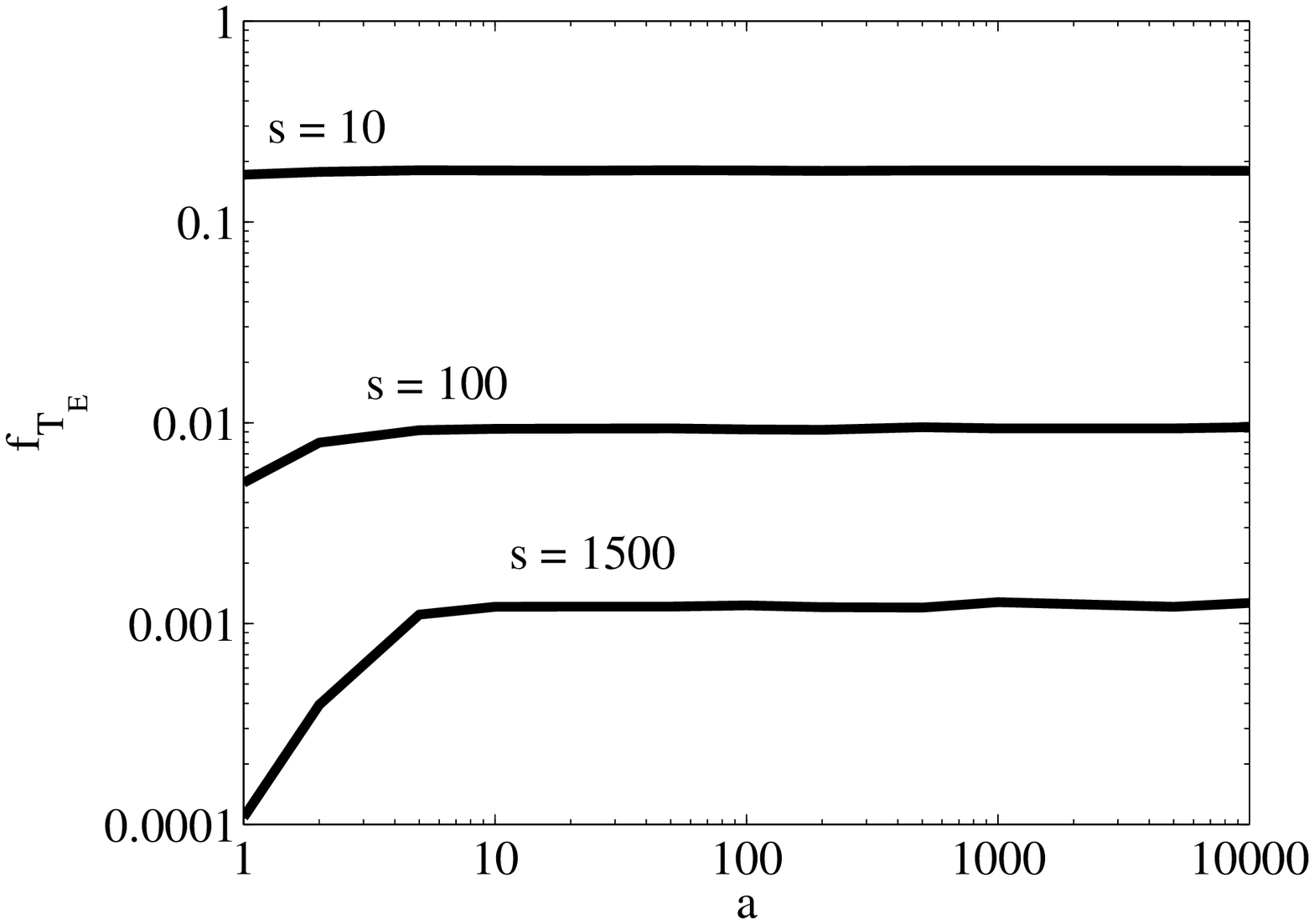}
\label{WallETimeFig}
}
\subfloat[]{
\includegraphics[scale = 0.55]{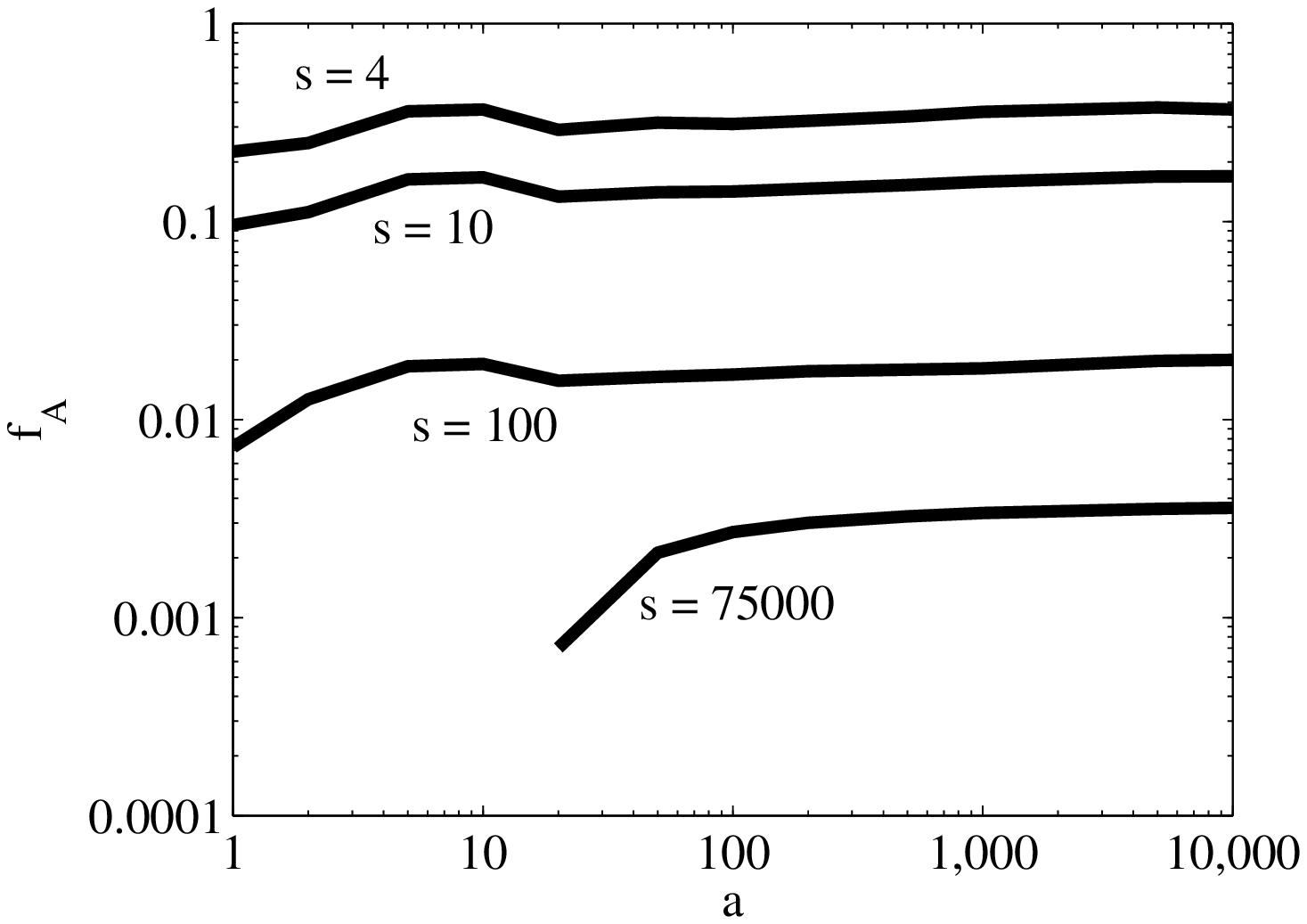}
\label{WallAreaFig}
}
\caption{Fraction of (a)~time squeezed $f_{T_E}$ and
(b)~\textit{area} squeezed $f_{A}$ vs.\ barrier position $a$ for
selected squeeze thresholds.  In~(a) we use a distribution of
trajectories of equally spaced energies, and moving out the barrier
does not lead to larger squeeze fractions.  As can be seen by
comparing~(b) with Fig.~\ref{WallTrajFigCS}, the area fraction
greater than some value is basically constant, modulo edge effects
near the barrier.}
\end{figure*}

The previous distribution was chosen with some amount of
arbitrariness. We could choose energies that fill
phase space out to $E_{min}$, the minimum allowed energy for
the given barrier position, and smooth out the bias towards higher
energy trajectories by weighting energies that are closer together
less than those that are farther apart.  Numerically this corresponds
to weighting each $f_{T_i}$ by
$\Delta E_{\textrm{traj}}/\Delta E_{\textrm{Tot}}$, where   
\begin{equation}
\Delta E_{\textrm{traj}} \equiv
  \lvert(E_{\textrm{next}}-E_{\textrm{previous}})\rvert/2 \label{dewt}
\end{equation}
and $\Delta E_{\textrm{Tot}}$ is the total energy range of the family of
trajectories from $0$ to $E_{\textrm{min}}$.
Thus we have  
\begin{equation}
f_{T_E} \equiv
  \sum {\frac{\Delta E_{\textrm{traj}}}{\Delta E_{\textrm{Tot}}}f_{T_i}}.
\label{sqewt}
\end{equation}
The results in this case are shown in Fig.~\ref{WallETimeFig} for
several different squeeze thresholds.  Now, as we move
out the barrier to larger values, the fraction of time spent squeezed
stays roughly constant (and a very small constant at that). Evaluating
the squeeze fraction with this measure does not support the intuition
presented in Sect.~\ref{Sec:USHOinfinite}.

We have therefore presented two different distributions for the
initial trajectories that show different behaviors in the limit as
the barrier is moved out.  This illustrates that the possibility of
having an increasing fraction of time squeezed is dependent on the
selection of the initial distribution of trajectories.

We reiterate the message of Fig.~\ref{TwoSizeTrajFig}: As the
barrier is moved to larger $q$ values, the minimum of the potential
goes down, opening up additional (lower) energy values that were
simply not available to the system when the barrier was closer.  
The two distributions we have
used so far treat these lower energy trajectories differently.  The
measure described by Eqn.~\ref{sqtraj} fixes the energy range to
one that is allowed for a close barrier position, and simply excludes
lower energies that become allowed as the barrier moves out.  The
measure described by Eqn.~\ref{sqewt} brings in more energies as they
become available (with a weighting that might be thought of as
``equipartition'' in energy). These differences are reflected in the
different outcomes regarding the preference (or otherwise) for
squeezing. 

\par
We now consider another measure that might be called ``equipartition in phase
space'': We simply measure a squeeze fraction weighted by phase space area.
We lay down a grid on our region 
of phase space and find the average value for $s$ in each box. This
means that the fraction of the area that is found above a given
squeeze threshold is simply
\begin{equation}
f_{A} \equiv \frac{N_{s >s_{\textrm{thresh}}}A_{\textrm{box}}}{A_{\textrm{Tot}}},
\label{sqarea}
\end{equation}
where $N$ is the number of boxes above a particular threshold,
$A_{\textrm{box}}$ is the area of the individual boxes, and
$A_{\textrm{Tot}}$ is the
total area of phase space. For the selected squeeze thresholds, we see
in Fig.~\ref{WallAreaFig} that roughly one percent of the area is
squeezed for a threshold of $100$ even as the barrier for
this system is moved to larger values of $a$. Larger thresholds exhibit even
less area squeezed. This supports the
conclusion that the amount of squeezing is constant as the size of the
system increases. 
In fact, inspection of the locations of the constant $s$ lines in
Fig.~\ref{WallTrajFigCS} makes it clear that a constant fraction of
the phase space lies above any fixed squeeze value, modulo ``edge
effects'' related to our discrete methodology, which are the source of
the the non-constant aspects of the curves in
Fig.~\ref{WallAreaFig}.

\par
After analyzing the upside-down simple harmonic oscillator with a
barrier, it appears that whether this finite system is overwhelmingly
squeezed or not depends on how the question is asked. In particular,
it appears that a bias or ``prior'' must be imposed on the
trajectories as the limit of distant barriers is considered
in order to realize the ``universal squeezing'' behavior discussed at
an intuitive level in Sect.~\ref{Sec:USHOinfinite}.  Thus if we
consider the distant barrier limit of the finite system to be a more
rigorous definition of the infinite case, the squeezing property is
not really a universal property of the system at all.  The heuristic
discussion from Sect.~\ref{Sec:USHOinfinite} is only recovered when
the limit is taken in a particular way, a way that imposes conditions
on what ``initial'' states are considered. 

\section{The Bent Phase Space Model}
\label{Sec:Bent}
\par
We have seen how trying to formulate the infinite USHO more rigorously
as the limit of a finite system undermined the rather appealing
intiuitive discussion we started with. But, perhaps this result is an
artifact of that particular toy model. 

\begin{figure}
\includegraphics[scale = 0.55]{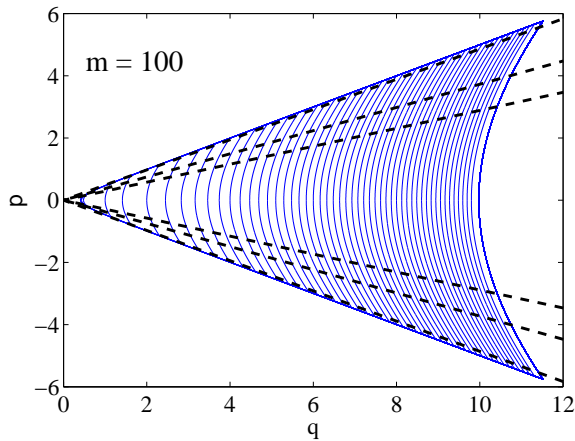}
\caption{Trajectories for the Bent Phase Space (BPS) model with $m = 100$ in
  the quadrant of phase space where $E < 0$ and $q > 0$.  
Constant $s_B$ curves are shown dashed, with values $s_B=70$, $7$ and
$4$. The BPS model was chosen to be more favorable to squeezing. 
 \label{TevTrajFig}}
\end{figure}

With that question in mind, we choose to look at squeezing for another model
that might show the
intended results that increasing the size of the system more generically
increases the
squeeze fraction. Rather than making the system finite by placing an
ordinary barrier at $q = a$, we choose a more complicated potential
that makes the phase space finite, such as letting $V$ have both $p$-
and $q$-dependence:  
\begin{equation}
V(p,q) =
 -\frac{q^2}{2}+\left(\frac{1}{2(m-(q^2-p^2))}\right)^{50}
 +\left(\frac{1}{10(q^2-4p^2)}\right)^{5}.
\label{BPSv}
\end{equation}
Changing the parameter $m$ corresponds to changing
the size of the system. Fig.~\ref{TevTrajFig} shows trajectories for
this model with $m = 100$. We call this the ``Bent Phase Space'' (BPS)
model. One other difference between this model
and the USHO model is that in this model the lines of squeezing are
now 
\begin{equation}p = \frac{q}{2}; p = -\frac{q}{2}.
\label{BPSasymp}
\end{equation} 
Since we have different
asymptotic lines of squeezing we define a different squeeze parameter:
\begin{equation}
s_B \equiv \left\lvert\frac{2p+q}{2p-q}\right\rvert
 + \left\lvert\frac{2p-q}{2p+q}\right\rvert. \label{sqtev}
\end{equation} 
This model is chosen because the squeezing appears more pronounced,
and because the barrier has a shape that appears to
fit the lines of squeezing better than a simple barrier at $q = a$. 

\par
\begin{figure*}
\includegraphics[scale = 0.6]{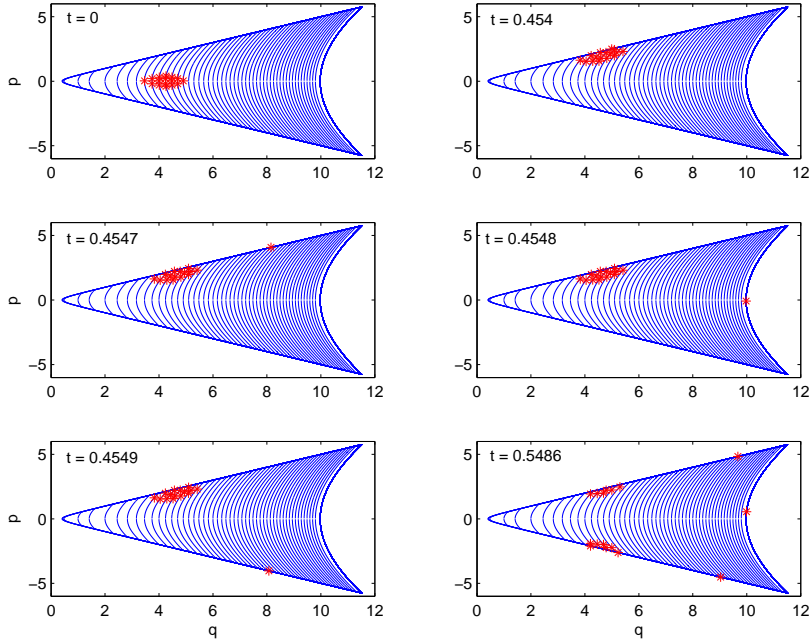}
\caption{Short squeeze duration in the BPS model: A patch started at $t = 0$
(as shown in upper-left) will spend
most of its time away from squeezed and barrier regions. As the
squeezed region is approached small parts of the patch will move
rapidly through to the other side. The time marked on each panel shows
the different time scales involved.  The patch is represented by a finite collection of
points. A more continuous treatment would show the patch stretched very
thinly along the trajectories followed by the isolated points. 
\label{BlobGridFig}}
\end{figure*}

We are able to continue with the same analysis as for the USHOb and ask the
same questions regarding how much time this model spends squeezed as
well as how much of the phase space area is squeezed. Since the potential
is now dependent on $p$, we no longer have $\dot{q} = p$ and instead have
\begin{equation}
\dot{q} =
p\left(1-\frac{100}{2^{50}}\frac{1}{(m-(q^2-p^2))^{51}}
 +\frac{4}{10^4}\frac{1}{(q^2-4p^2)^6}\right).
\label{BPSqdot}
\end{equation}
One can see that close to the asymptotes (given by
Eqn.~\ref{BPSasymp}) the third term in $\dot{q}$ diverges. This means
that a particle at a given energy will spend most of its time far away
from the barrier and will pass close to asymptotes very quickly.  It
will also bounce off the barrier very quickly since
$\dot{q}$ is large in that region. Fig.~\ref{BlobGridFig} shows this 
for a particular patch. One can also understand this effect
intuitively by noting how the trajectories for the BPS model pack
extremely closely together near the asymptotes.  Liouville's theorem
then tells us that the patch must spread out extremely broadly in the
direction along the trajectories. 

\par
Once again the relevant concern is what happens as the size of the system
grows large. The shape of the barrier seems to respect squeezing better than
the USHOb model,
yet a given particle moves through the squeezed region very quickly. To
determine the net effect we choose a fixed squeeze threshold, $s_B$
and ask what happens to the squeeze fraction as we increase the parameter $m$.   

\par
\begin{figure}
\includegraphics[scale = 0.45]{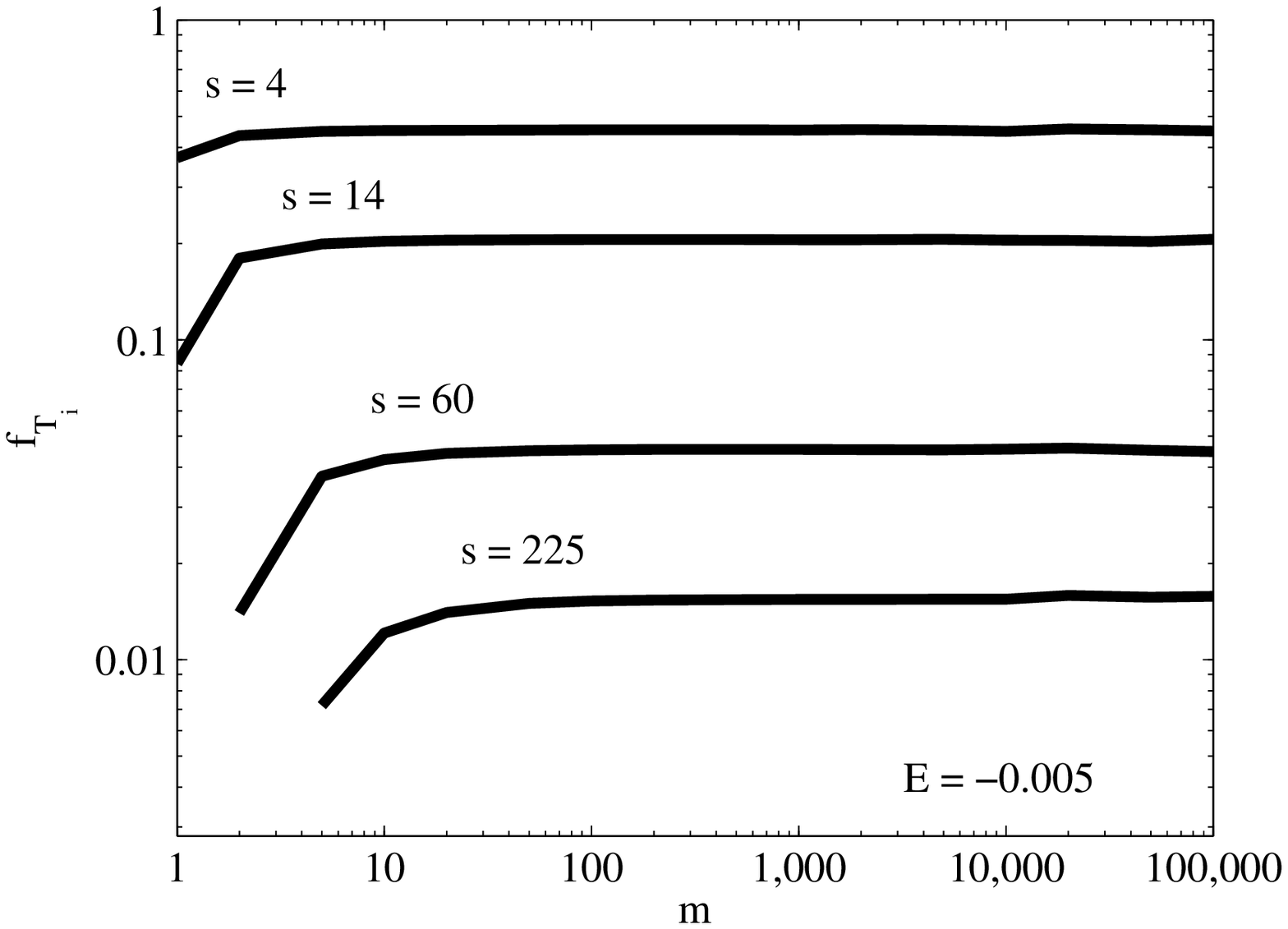}
\caption{The BPS model: Fraction of time squeezed $f_{T_{i}}$
  vs. barrier position $m$ for selected
  squeeze thresholds for an energy, $E = -0.005$, favorable to
  squeezing. In contrast to Fig.~\ref{WallE1e3TimeFig} for the USHOb,
  the amount of squeezing does not increase with $m$, even for this
  very favorable case. The rapid motion of the BPS trajectories
  through the squeezed region is the reason that squeezing is not
  strongly favored in this model.\label{TevE5e3TimeFig}}
\end{figure}

As before, we start by choosing a larger energy trajectory since we
expect this will spend a large amount of time squeezed. Again, we
arbitrarily choose several different squeeze thresholds. Using the
same definition for $f_{T_i}$, we see in Fig.~\ref{TevE5e3TimeFig} that
the fraction of time spent squeezed remains constant even as the size of the
system increases. The expected effect due to the increase in size is canceled 
out by the increased speed of a particle in the squeezed regions. Very large
squeeze values
spend a negligible amount of time squeezed even for a favorable energy such as
the one shown. Rather than explicitly considering the other ``measures''
defined for the USHOb model (specifically Eqns.~\ref{sqtraj} and \ref{sqewt}),
we see that since even an energy that is favorable toward squeezing shows a
constant amount of time squeezed, any distribution of trajectories will
display the same effects. The only difference between any of these time-based
``measures'' will be the exact value of the small
fraction of time the distribution will spend squeezed. 

\par
\begin{figure}
\includegraphics[scale = 0.55]{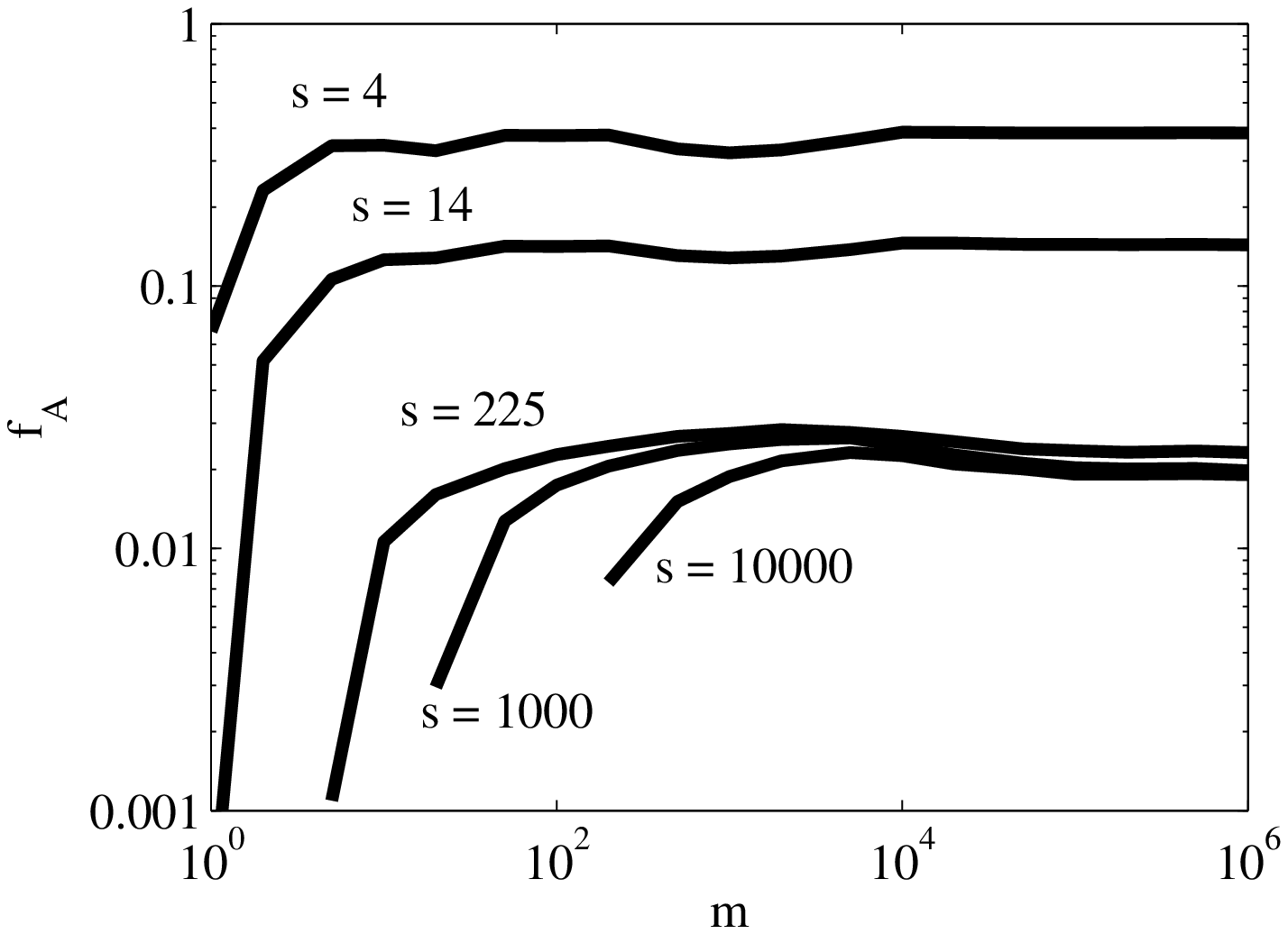}
\caption{The BPS model: Fraction of \textit{area} squeezed $f_{A}$
  vs. barrier position $m$ for
  selected squeeze thresholds. As can be seen from Fig
  \ref{TevTrajFig} a constant area independent of $m$ is the natural
  large $m$ behavior.
  \hfill\break 
\label{TevAreaFig}}
\end{figure}

Finally, we look at the fraction of phase space area squeezed for this model
(Fig.~\ref{TevAreaFig}). The fraction of the area of phase space that is
squeezed asymptotes to a constant value as well. This is clear from the fact
that curves of constant squeezing in this model are once again straight lines
like the USHOb model. Overall, it appears that rather than improving the
USHOb model, the BPS model is even more unfavorable to squeezing. 

\section{Interpretation and Conclusions}
\label{Sec:Conclusions}

\subsection{The Toy Models}
\label{Sec:ConToy}
\par
In Section \ref{Sec:USHOinfinite} we provided an intuitive analysis
of the upside-down harmonic oscillator (USHO) which suggested that it
had a highly universal squeezing behavior, independent of the initial
conditions. Specifically, any initial region in phase space would in
due course spend all of eternity in an arbitrarily squeezed state. As
we discussed in the introduction, it has been hoped 
that this kind of behavior illustrates a phenomenon that could be put
to good use in tackling difficult problems facing theoretical
cosmology \cite{Guth:2011aa}.  However the USHO is an infinite system, and our heuristic
discussion was quite loose in the way the infinity was handled. This
paper has sought to study the phenomenon more rigorously by considering
infinite models as limits of finite systems as their sizes are taken
to infinity. 

We have quantitatively analyzed the amount of squeezing for two 
different finite models. The first model was constructed by placing a barrier
at $q = a$ for the USHO (creating the ``USHOb'' model).  This
model showed an increasing fraction of time squeezed as the 
size of the system increased for a specific chosen trajectory, as
well as for a distribution of trajectories biased toward higher
energies so as to favored squeezing. However, this model also showed squeeze
fractions
that were constant in time for a distribution that eliminated the
higher energy bias.  Thus for the USHOb model special assumptions or
biases about the state of the system (i.e.~the energies of the
trajectories) are needed to recover the ``eternal'' squeezing behavior in the
large
system limit. If the limiting behavior of this system is taken to be a
more rigorous description of the USHO model the attractive features of
the USHO model do not look universal after all.  Instead they appear
to be closely tied to special assumptions about the state of the system. 

\par
The second model (the BPS model) had a more complicated behavior which
superficially seemed more favorable to 
squeezing. However, when analyzed systematically this model too
could not describe the hoped-for universal squeezing behavior.  In
fact, it's behavior was even more problematic in that regard. 

\par
For each model, the squeeze values analyzed were chosen
arbitrarily. The squeeze fractions were plotted versus the size of the
system. However, another way of showing the same conclusions for each
model was by plotting the squeeze fraction versus squeeze value and
placing multiple system sizes on the same plot. Using this method, we
are able to see trends in the squeeze values on a larger scale at the
cost of being able to see the squeezing for a given value less
clearly. These alternative methods of analysis
support the conclusions previously stated in this paper.  

Of course, all we have done is consider limits of two specific finite
models.  Perhaps there are other models that would offer a more positive
outcome. And as we have discussed at the beginning, how the behavior of such
toy models can be used to give insights into cosmological theories is
a poorly developed subject.  While we do not claim to have advanced
that subject here, we conclude with some general comments about
related issues in cosmology.

\subsection{Connection to Inflation}
\label{Sec:ConInf}

The project of calculating the probability of inflation taking place
has generated considerable interest and controversy (for example~\cite{Guth:2007ng,Hartle:1983ai,*Gibbons:1986xk,*Hawking:1987bi,*Kachru:2003aw,*Gibbons:2006pa,*Linde:2007fr,*Carroll:2010aj,Schiffrin:2012zf}). 
People have tried a variety of approaches from classical phase space
arguments to ``wavefunctions of the universe'' in quantum cosmology,
to the dynamics of the string theory landscape,
but so far there is no agreement as to the ``right answer''.
As emphasized in \cite{Albrecht:2004ke}, the probability for inflation
to start is an essential ingredient of any claim that inflation describes the
most likely way of creating our observed universe. 
None the less, it is frequently hoped that in the limit where eternal inflation
really lasts forever one can neglect any concerns about how
low the probability may be of starting eternal 
inflation in the first place.  Surely the infinite volume produced by eternal
inflation will outweigh even an arbitrarily low probability for it to start,
as long as that probability is nonzero.  This would seem to allow us
to make predictions despite the uncertainties surrounding the start of
inflation. 

More specifically, in Guth's discussion of his toy model
he argues that the squeezing (which he calls ``tuning'' in reference
to the tuning of ``initial'' conditions required for realistic pocket universes) is completely
generic \cite{Guth:2011aa} regardless of initial conditions. He then
argues that the infinite phase space assumed in models of eternal
inflation could lead to inflation and being generic through a similar
sort of behavior.   The intuitive picture of the
USHO presented in Section \ref{Sec:USHOinfinite} is just a restatement
of this original argument by Guth.  It appears to be a way that the
dynamical behavior of a system can create a universal outcome
regardless of initial conditions. 

But to realize the hoped-for picture of eternal inflation one really needs to
be much more rigorous than anyone has been so far. Schiffrin and
Wald~\cite{Schiffrin:2012zf} review some of these issues in a manner
very much in the spirit of this paper.  They identify not only the
lack of equilibrium behavior emphasized and exploited by Guth
\cite{Guth:2011aa}, but also the importance of defining infinite
systems rigorously, which is the focus of this paper.
In the case of the toy model, we have attempted to improve the rigor by considering
infinite systems as limiting cases of finite ones. This model can be thought of as a
concrete realization of the limiting process discussed formally in
section IV of~\cite{Schiffrin:2012zf}. Our approach should not be confused with
various cutoffs used to regulate the infinities that come from the
pocket universe counting problem. For one, imposing these cutoffs does
not result in Hamiltonian systems, whereas all our models are
Hamiltonian.  Secondly, pocket universe formation is typically a quantum process
which is clearly not present in our classical toy models. Also, as originally conceived in~\cite{Guth:2011aa}
the toy model was not intended to address the pocket universe counting problem.  While we
would be curious if the toy models offer an interesting perspective on
that problem, we focus on Guth's original topic of interest in this paper.

The failure of our
finite toy models to realize the desired universal squeezing behavior
in large size limit is an illustration
of how things can go wrong.  We add that the one case we know where
eternal inflation has been considered as limit of a well-regulated
finite system the outcome was opposite to the usual
beliefs: The probability of starting inflation decreased sufficiently
rapidly with increasing inflation duration that even when maximum
impact of the increasing volume was assumed the probability of us
living in an eternally inflating universe dropped to zero as the
infinite limit was taken \cite{Albrecht:2004ke}. 

We have investigated simple toy models that we hoped would exhibit universal
dynamics of the
sort that would be nice to utilize in cosmological theories.  Our
attempts here to more rigorously treat earlier ideas along these lines
yielded discouraging results, in that the desired universal
behavior proved not so universal after all.  We hope this work will stimulate
other explorations which perhaps
could yield toy models that do better.  However, we suspect that the
absence of successful models so far probably reflects a deeper
problem with the idea that some sort of universal cosmological
dynamics could explain the ``typicality'' of the universe we observe. 

\acknowledgements
We thank A. Guth and D. Phillips for helpful conversations.  
This collaboration grew out of conversations at the Third 
International FQXi conference in 2011; AA and TD are grateful 
to FQXi for supporting their participation. AA thanks the KICP and the 
University of Chicago department of Astronomy and Astrophysics for
hospitality during his sabbatical. This work was supported in part by
DOE Grant DE-FG03-91ER40674 and UC Davis.  

\bibliography{AA}

\begin{thebibliography}{19}%
\makeatletter
\providecommand \@ifxundefined [1]{%
 \@ifx{#1\undefined}
}%
\providecommand \@ifnum [1]{%
 \ifnum #1\expandafter \@firstoftwo
 \else \expandafter \@secondoftwo
 \fi
}%
\providecommand \@ifx [1]{%
 \ifx #1\expandafter \@firstoftwo
 \else \expandafter \@secondoftwo
 \fi
}%
\providecommand \natexlab [1]{#1}%
\providecommand \enquote  [1]{``#1''}%
\providecommand \bibnamefont  [1]{#1}%
\providecommand \bibfnamefont [1]{#1}%
\providecommand \citenamefont [1]{#1}%
\providecommand \href@noop [0]{\@secondoftwo}%
\providecommand \href [0]{\begingroup \@sanitize@url \@href}%
\providecommand \@href[1]{\@@startlink{#1}\@@href}%
\providecommand \@@href[1]{\endgroup#1\@@endlink}%
\providecommand \@sanitize@url [0]{\catcode `\\12\catcode `\$12\catcode
  `\&12\catcode `\#12\catcode `\^12\catcode `\_12\catcode `\%12\relax}%
\providecommand \@@startlink[1]{}%
\providecommand \@@endlink[0]{}%
\providecommand \url  [0]{\begingroup\@sanitize@url \@url }%
\providecommand \@url [1]{\endgroup\@href {#1}{\urlprefix }}%
\providecommand \urlprefix  [0]{URL }%
\providecommand \Eprint [0]{\href }%
\providecommand \doibase [0]{http://dx.doi.org/}%
\providecommand \selectlanguage [0]{\@gobble}%
\providecommand \bibinfo  [0]{\@secondoftwo}%
\providecommand \bibfield  [0]{\@secondoftwo}%
\providecommand \translation [1]{[#1]}%
\providecommand \BibitemOpen [0]{}%
\providecommand \bibitemStop [0]{}%
\providecommand \bibitemNoStop [0]{.\EOS\space}%
\providecommand \EOS [0]{\spacefactor3000\relax}%
\providecommand \BibitemShut  [1]{\csname bibitem#1\endcsname}%
\let\auto@bib@innerbib\@empty
\bibitem [{\citenamefont {Starobinsky}(1980)}]{Starobinsky:1980te}%
  \BibitemOpen
  \bibfield  {author} {\bibinfo {author} {\bibfnamefont {A.~A.}\ \bibnamefont
  {Starobinsky}},\ }\href {\doibase 10.1016/0370-2693(80)90670-X} {\bibfield
  {journal} {\bibinfo  {journal} {Phys.Lett.}\ }\textbf {\bibinfo {volume}
  {B91}},\ \bibinfo {pages} {99} (\bibinfo {year} {1980})}\BibitemShut
  {NoStop}%
\bibitem [{\citenamefont {Guth}(1981)}]{Guth:1980zm}%
  \BibitemOpen
  \bibfield  {author} {\bibinfo {author} {\bibfnamefont {A.~H.}\ \bibnamefont
  {Guth}},\ }\href {\doibase 10.1103/PhysRevD.23.347} {\bibfield  {journal}
  {\bibinfo  {journal} {Phys.Rev.}\ }\textbf {\bibinfo {volume} {D23}},\
  \bibinfo {pages} {347} (\bibinfo {year} {1981})}\BibitemShut {NoStop}%
\bibitem [{\citenamefont {Linde}(1982)}]{Linde:1981mu}%
  \BibitemOpen
  \bibfield  {author} {\bibinfo {author} {\bibfnamefont {A.~D.}\ \bibnamefont
  {Linde}},\ }\href {\doibase 10.1016/0370-2693(82)91219-9} {\bibfield
  {journal} {\bibinfo  {journal} {Phys.Lett.}\ }\textbf {\bibinfo {volume}
  {B108}},\ \bibinfo {pages} {389} (\bibinfo {year} {1982})}\BibitemShut
  {NoStop}%
\bibitem [{\citenamefont {Albrecht}\ and\ \citenamefont
  {Steinhardt}(1982)}]{Albrecht:1982wi}%
  \BibitemOpen
  \bibfield  {author} {\bibinfo {author} {\bibfnamefont {A.}~\bibnamefont
  {Albrecht}}\ and\ \bibinfo {author} {\bibfnamefont {P.~J.}\ \bibnamefont
  {Steinhardt}},\ }\href {\doibase 10.1103/PhysRevLett.48.1220} {\bibfield
  {journal} {\bibinfo  {journal} {Phys.Rev.Lett.}\ }\textbf {\bibinfo {volume}
  {48}},\ \bibinfo {pages} {1220} (\bibinfo {year} {1982})}\BibitemShut
  {NoStop}%
\bibitem [{\citenamefont {Guth}(2007)}]{Guth:2007ng}%
  \BibitemOpen
  \bibfield  {author} {\bibinfo {author} {\bibfnamefont {A.~H.}\ \bibnamefont
  {Guth}},\ }\href {\doibase 10.1088/1751-8113/40/25/S25} {\bibfield  {journal}
  {\bibinfo  {journal} {J. Phys.}\ }\textbf {\bibinfo {volume} {A40}},\
  \bibinfo {pages} {6811} (\bibinfo {year} {2007})}\BibitemShut {NoStop}%
\bibitem [{\citenamefont {Albrecht}\ and\ \citenamefont
  {Phillips}(2012)}]{Albrecht:2012zp}%
  \BibitemOpen
  \bibfield  {author} {\bibinfo {author} {\bibfnamefont {A.}~\bibnamefont
  {Albrecht}}\ and\ \bibinfo {author} {\bibfnamefont {D.}~\bibnamefont
  {Phillips}},\ }\href@noop {} {\  (\bibinfo {year} {2012})},\ \Eprint
  {http://arxiv.org/abs/1212.0953} {arXiv:1212.0953 [gr-qc]} \BibitemShut
  {NoStop}%
\bibitem [{\citenamefont {Albrecht}(2009)}]{Albrecht:2009vr}%
  \BibitemOpen
  \bibfield  {author} {\bibinfo {author} {\bibfnamefont {A.}~\bibnamefont
  {Albrecht}},\ }\href {\doibase 10.1088/1742-6596/174/1/012006} {\bibfield
  {journal} {\bibinfo  {journal} {J. Phys. Conf. Ser.}\ }\textbf {\bibinfo
  {volume} {174}},\ \bibinfo {pages} {012006} (\bibinfo {year} {2009})},\
  \Eprint {http://arxiv.org/abs/0906.1047} {arXiv:0906.1047 [gr-qc]}
  \BibitemShut {NoStop}%
\bibitem [{\citenamefont {Albrecht}(2011)}]{Albrecht:2011yg}%
  \BibitemOpen
  \bibfield  {author} {\bibinfo {author} {\bibfnamefont {A.}~\bibnamefont
  {Albrecht}},\ }\href {\doibase 10.1103/PhysRevLett.107.151102} {\bibfield
  {journal} {\bibinfo  {journal} {Phys.Rev.Lett.}\ }\textbf {\bibinfo {volume}
  {107}},\ \bibinfo {pages} {151102} (\bibinfo {year} {2011})},\ \Eprint
  {http://arxiv.org/abs/1104.3315} {arXiv:1104.3315 [astro-ph.CO]} \BibitemShut
  {NoStop}%
\bibitem [{\citenamefont {Albrecht}(2012)}]{Albrecht:2012aa}%
  \BibitemOpen
  \bibfield  {author} {\bibinfo {author} {\bibfnamefont {A.}~\bibnamefont
  {Albrecht}},\ }\href@noop {} {\  (\bibinfo {year} {2012})},\ \bibinfo {note}
  {in Preparation}\BibitemShut {NoStop}%
\bibitem [{\citenamefont {Guth}(2011)}]{Guth:2011aa}%
  \BibitemOpen
  \bibfield  {author} {\bibinfo {author} {\bibfnamefont {A.}~\bibnamefont
  {Guth}},\ }\href@noop {} {\  (\bibinfo {year} {2011})},\ \bibinfo {note}
  {talk presented at the {\em Challenges for Early Universe Cosmology}
  conference, Perimeter Institute}\BibitemShut {NoStop}%
\bibitem [{\citenamefont {Hartle}\ and\ \citenamefont
  {Hawking}(1983)}]{Hartle:1983ai}%
  \BibitemOpen
  \bibfield  {author} {\bibinfo {author} {\bibfnamefont {J.~B.}\ \bibnamefont
  {Hartle}}\ and\ \bibinfo {author} {\bibfnamefont {S.~W.}\ \bibnamefont
  {Hawking}},\ }\href@noop {} {\bibfield  {journal} {\bibinfo  {journal} {Phys.
  Rev.}\ }\textbf {\bibinfo {volume} {D28}},\ \bibinfo {pages} {2960} (\bibinfo
  {year} {1983})}\BibitemShut {NoStop}%
\bibitem [{\citenamefont {Gibbons}\ \emph {et~al.}(1987)\citenamefont
  {Gibbons}, \citenamefont {Hawking},\ and\ \citenamefont
  {Stewart}}]{Gibbons:1986xk}%
  \BibitemOpen
  \bibfield  {author} {\bibinfo {author} {\bibfnamefont {G.}~\bibnamefont
  {Gibbons}}, \bibinfo {author} {\bibfnamefont {S.}~\bibnamefont {Hawking}}, \
  and\ \bibinfo {author} {\bibfnamefont {J.}~\bibnamefont {Stewart}},\ }\href
  {\doibase 10.1016/0550-3213(87)90425-1} {\bibfield  {journal} {\bibinfo
  {journal} {Nucl.Phys.}\ }\textbf {\bibinfo {volume} {B281}},\ \bibinfo
  {pages} {736} (\bibinfo {year} {1987})}\BibitemShut {NoStop}%
\bibitem [{\citenamefont {Hawking}\ and\ \citenamefont
  {Page}(1988)}]{Hawking:1987bi}%
  \BibitemOpen
  \bibfield  {author} {\bibinfo {author} {\bibfnamefont {S.}~\bibnamefont
  {Hawking}}\ and\ \bibinfo {author} {\bibfnamefont {D.~N.}\ \bibnamefont
  {Page}},\ }\href {\doibase 10.1016/0550-3213(88)90008-9} {\bibfield
  {journal} {\bibinfo  {journal} {Nucl.Phys.}\ }\textbf {\bibinfo {volume}
  {B298}},\ \bibinfo {pages} {789} (\bibinfo {year} {1988})}\BibitemShut
  {NoStop}%
\bibitem [{\citenamefont {Kachru}\ \emph {et~al.}(2003)\citenamefont {Kachru},
  \citenamefont {Kallosh}, \citenamefont {Linde},\ and\ \citenamefont
  {Trivedi}}]{Kachru:2003aw}%
  \BibitemOpen
  \bibfield  {author} {\bibinfo {author} {\bibfnamefont {S.}~\bibnamefont
  {Kachru}}, \bibinfo {author} {\bibfnamefont {R.}~\bibnamefont {Kallosh}},
  \bibinfo {author} {\bibfnamefont {A.~D.}\ \bibnamefont {Linde}}, \ and\
  \bibinfo {author} {\bibfnamefont {S.~P.}\ \bibnamefont {Trivedi}},\ }\href
  {\doibase 10.1103/PhysRevD.68.046005} {\bibfield  {journal} {\bibinfo
  {journal} {Phys.Rev.}\ }\textbf {\bibinfo {volume} {D68}},\ \bibinfo {pages}
  {046005} (\bibinfo {year} {2003})},\ \Eprint
  {http://arxiv.org/abs/hep-th/0301240} {arXiv:hep-th/0301240 [hep-th]}
  \BibitemShut {NoStop}%
\bibitem [{\citenamefont {Gibbons}\ and\ \citenamefont
  {Turok}(2008)}]{Gibbons:2006pa}%
  \BibitemOpen
  \bibfield  {author} {\bibinfo {author} {\bibfnamefont {G.}~\bibnamefont
  {Gibbons}}\ and\ \bibinfo {author} {\bibfnamefont {N.}~\bibnamefont
  {Turok}},\ }\href {\doibase 10.1103/PhysRevD.77.063516} {\bibfield  {journal}
  {\bibinfo  {journal} {Phys.Rev.}\ }\textbf {\bibinfo {volume} {D77}},\
  \bibinfo {pages} {063516} (\bibinfo {year} {2008})},\ \Eprint
  {http://arxiv.org/abs/hep-th/0609095} {arXiv:hep-th/0609095 [hep-th]}
  \BibitemShut {NoStop}%
\bibitem [{\citenamefont {Linde}(2008)}]{Linde:2007fr}%
  \BibitemOpen
  \bibfield  {author} {\bibinfo {author} {\bibfnamefont {A.}~\bibnamefont
  {Linde}},\ }\href {\doibase 10.1007/978-3-540-74353-8_1} {\bibfield
  {journal} {\bibinfo  {journal} {Lect. Notes Phys.}\ }\textbf {\bibinfo
  {volume} {738}},\ \bibinfo {pages} {1} (\bibinfo {year} {2008})},\ \Eprint
  {http://arxiv.org/abs/0705.0164} {arXiv:0705.0164 [hep-th]} \BibitemShut
  {NoStop}%
\bibitem [{\citenamefont {Carroll}\ and\ \citenamefont
  {Tam}(2010)}]{Carroll:2010aj}%
  \BibitemOpen
  \bibfield  {author} {\bibinfo {author} {\bibfnamefont {S.~M.}\ \bibnamefont
  {Carroll}}\ and\ \bibinfo {author} {\bibfnamefont {H.}~\bibnamefont {Tam}},\
  }\href@noop {} {\  (\bibinfo {year} {2010})},\ \Eprint
  {http://arxiv.org/abs/1007.1417} {arXiv:1007.1417 [hep-th]} \BibitemShut
  {NoStop}%
\bibitem [{\citenamefont {Schiffrin}\ and\ \citenamefont
  {Wald}(2012)}]{Schiffrin:2012zf}%
  \BibitemOpen
  \bibfield  {author} {\bibinfo {author} {\bibfnamefont {J.~S.}\ \bibnamefont
  {Schiffrin}}\ and\ \bibinfo {author} {\bibfnamefont {R.~M.}\ \bibnamefont
  {Wald}},\ }\href {\doibase 10.1103/PhysRevD.86.023521} {\bibfield  {journal}
  {\bibinfo  {journal} {Phys.Rev.}\ }\textbf {\bibinfo {volume} {D86}},\
  \bibinfo {pages} {023521} (\bibinfo {year} {2012})},\ \Eprint
  {http://arxiv.org/abs/1202.1818} {arXiv:1202.1818 [gr-qc]} \BibitemShut
  {NoStop}%
\bibitem [{\citenamefont {Albrecht}\ and\ \citenamefont
  {Sorbo}(2004)}]{Albrecht:2004ke}%
  \BibitemOpen
  \bibfield  {author} {\bibinfo {author} {\bibfnamefont {A.}~\bibnamefont
  {Albrecht}}\ and\ \bibinfo {author} {\bibfnamefont {L.}~\bibnamefont
  {Sorbo}},\ }\href@noop {} {\bibfield  {journal} {\bibinfo  {journal} {Phys.
  Rev.}\ }\textbf {\bibinfo {volume} {D70}},\ \bibinfo {pages} {063528}
  (\bibinfo {year} {2004})},\ \Eprint {http://arxiv.org/abs/hep-th/0405270}
  {hep-th/0405270} \BibitemShut {NoStop}%
\end{thebibliography}%

\end{document}